\documentclass[twocolumn,preprintnumbers,amsmath,amssymb]{revtex4}

\usepackage{graphicx}
\usepackage{dcolumn}
\usepackage{bm}

\begin{document}

\title{Stress-induced modification of the boson peak scaling behavior}

\author{S. Corezzi$^{1,\footnotemark[1],\footnotemark[2]}$, S. Caponi$^{2,3,\footnotemark[2]}$, F. Rossi$^{2}$, D. Fioretto$^{1,4}$}
\affiliation{$^1$Dipartimento di Fisica, Universit$\grave{a}$ di Perugia, Via A. Pascoli, I-06123 Perugia, Italy\\
$^2$Dipartimento di Fisica, Universit\`{a} di Trento, Via Sommarive 14, 38050 Povo (Trento), Italy\\
$^3$Istituto di Biofisica, Consiglio Nazionale delle Ricerche, Via alla Cascata 56/C, 38123 Trento, Italy\\
$^4$Centro di Eccellenza sui Materiali Innovativi Nanostrutturati
(CEMIN), Universit\`{a} di Perugia, Via Elce di Sotto 8, 06123
Perugia, Italy}

\date{June 3, 2013}

\begin{abstract}
The scaling behavior of the so-called boson peak in glass-formers
and its relation to the elastic properties of the system remains a
source of controversy. Here, the boson peak in a binary reactive
mixture is measured by Raman scattering (i) on cooling the unreacted
mixture well below its glass transition temperature and (ii) after
quenching to very low temperature the mixture at different times
during isothermal polymerization. These different paths to the
glassy phase are able to generate glasses with different amounts of
residual stresses, as evidenced by the departure of the elastic
moduli from a Cauchy-like relationship. We find that the scaling
behavior of the boson peak with the properties of the elastic medium
--- as measured by the Debye frequency --- holds for states in which
the system is able to release internal stress and breaks down in the
presence of residual stresses. These findings provide new insight
into the boson peak behavior and are able to reconcile the
apparently conflicting results presented in literature.

\end{abstract}


\begin{figure}
\includegraphics[width=6.5 cm]{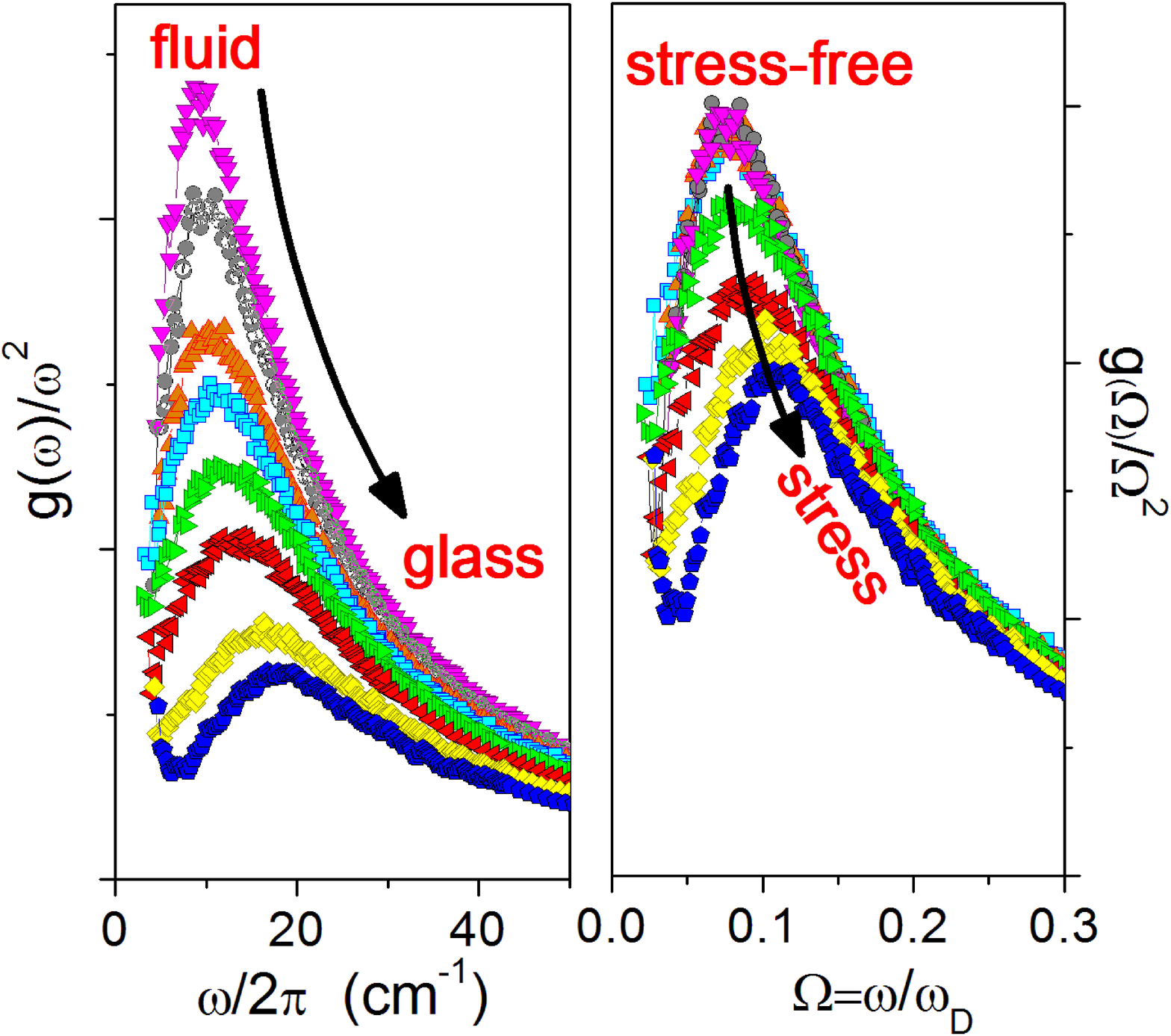}
\end{figure}

 \maketitle

\footnotetext[1]{* Corresponding author:
silvia.corezzi@fisica.unipg.it}

\footnotetext[2]{\dag ~ S.C. and S.C. contributed equally.}

\section{Introduction}

It is now well established that the density of vibrational states,
$g(\omega)$, of amorphous materials differs in a characteristic way
from that of crystalline ones. Whereas the low frequency $g(\omega)$
in crystals is well predicted by the Debye model, the lack of
long-range periodicity is responsible for the universal presence, in
amorphous materials, of an excess contribution over the Debye
prediction, evidenced as a peak in the reduced
$g(\omega)/\omega^{2}$ --- the so-called boson peak (BP). This
anomalous feature appears in the signal detected by several
techniques, such as inelastic neutron scattering
\cite{FrickSCIENCE1995}, calorimetry \cite{ZellerPRB1971}, nuclear
inelastic scattering \cite{MonacoPRL2006hyp, MonacoPRL2006den},
M\"{o}ssbauer \cite{AchterholdPRE2002}, Raman \cite{SokolovPRB1995}
and hyper-Raman spectroscopy \cite{SimonPRL2006, HehlenJRS2012}.
Despite the long-standing interest in this characteristic feature,
its origin still remains a source of controversy
\cite{ParshinPRB2007, SchirmacherPRL2007, TanakaNATURE2008,
RuoccoSCIREP2013, Zanatta2011, Baldi2009, Ruffle2010, Chumakov2011}.

An approach that conveys information without invoking any model is
studying the changes of these excess vibrational modes under the
application of external stimuli such as temperature and pressure, or
by varying the sample conditions such as the volume available to the
system, the thermal history or the number of bonds between the
molecules. Without exception, available results are all in agreement
\cite{Zanatta2011, Baldi2009, Ruffle2010, Chumakov2011, Zanatta2010,
RossiJPCB2012, CaponiPRL2009, MonacoPRL2006den, NissPRL2007,
MonacoPRL2006hyp, Hong2008, CaponiJPCM2007}: the BP shifts towards
higher frequencies and decreases in intensity when the sample
becomes stiffer (increase of elastic constants). Recently, it has
been proposed to quantitatively test the role played by the changes
in the elastic properties by scaling the BP data with the Debye
frequency,  a quantity defined by the elastic medium. The question
of whether such a scaling is able to generate a master curve of the
data, and hence, the variations of $g(\omega)$ can be explained in
terms of elastic medium transformations is still highly debated. So
far, the accumulated evidence is apparently in favor of opposite
views: on one hand, the BP evolution in sodium silicate glasses
(cooled \cite{Baldi2009}, compressed \cite{Chumakov2011},
hyperquenched \cite{MonacoPRL2006hyp} and permanently densified
samples \cite{MonacoPRL2006den}) and in an epoxy-amine mixture
during chemical vitrification \cite{CaponiPRL2009} is controlled by
changes in the system's elasticity, according to the Debye scaling
law; on the other hand, the BP variations in network forming glasses
like silica and GeO$_{2}$ upon cooling \cite{Ruffle2010,
Zanatta2011, CaponiPRB2007}, in permanently densified silica
\cite{Zanatta2010}, and in few polymers under pressure
\cite{NissPRL2007, Hong2008} are stronger than the elastic medium
transformation, and the Debye scaling does not work. In light of
such controversial results the question we face is the following:
\emph{Do these results really conflict, or rather there is a general
explanation in terms of the elastic properties of the systems?}

\begin{figure*}
\vspace{-0.8 cm}
\includegraphics[width=7.5 cm]{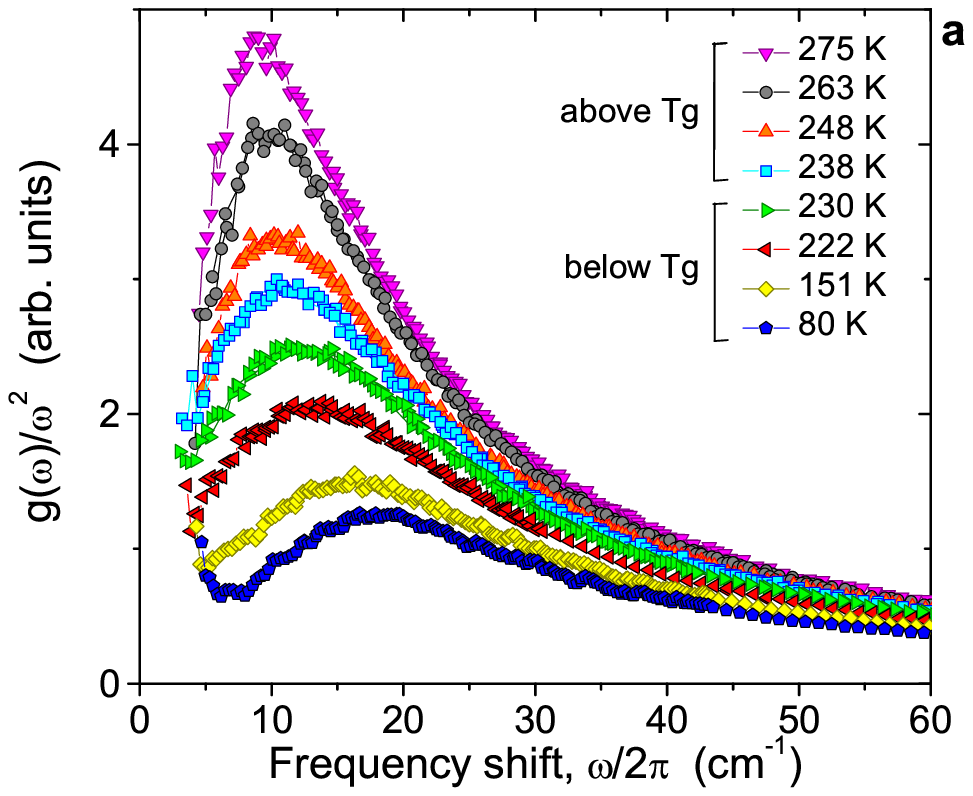}
\includegraphics[width=7.5 cm]{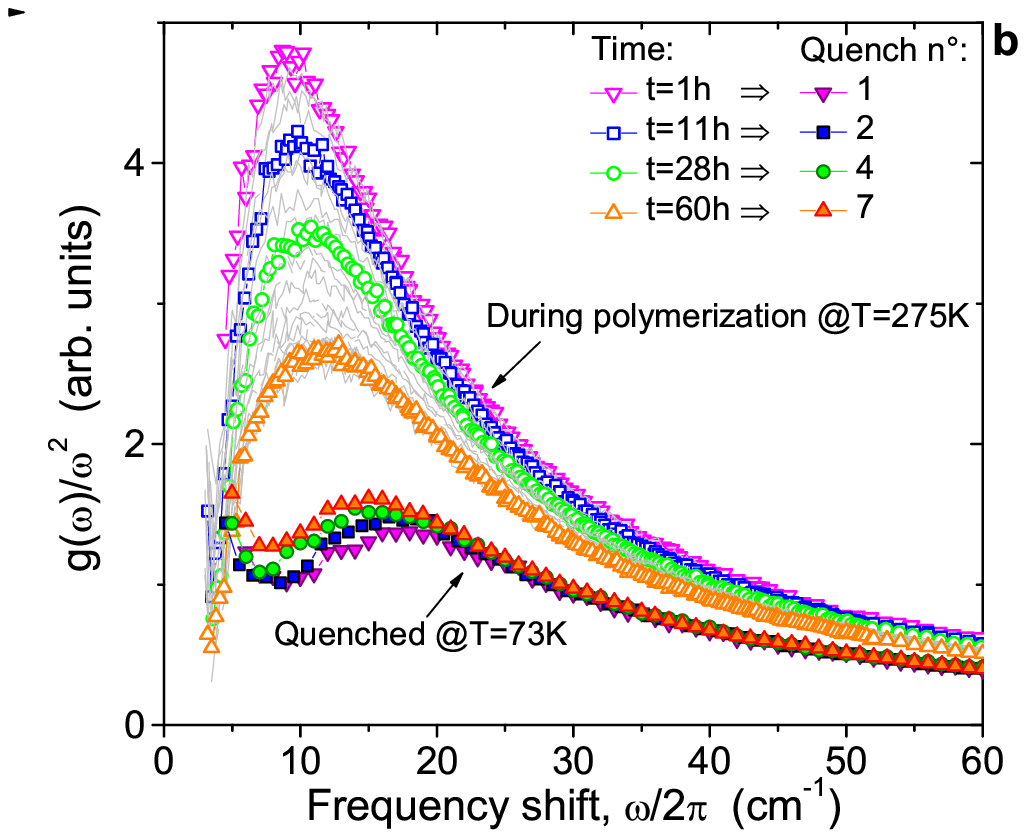}
\includegraphics[width=7.5 cm]{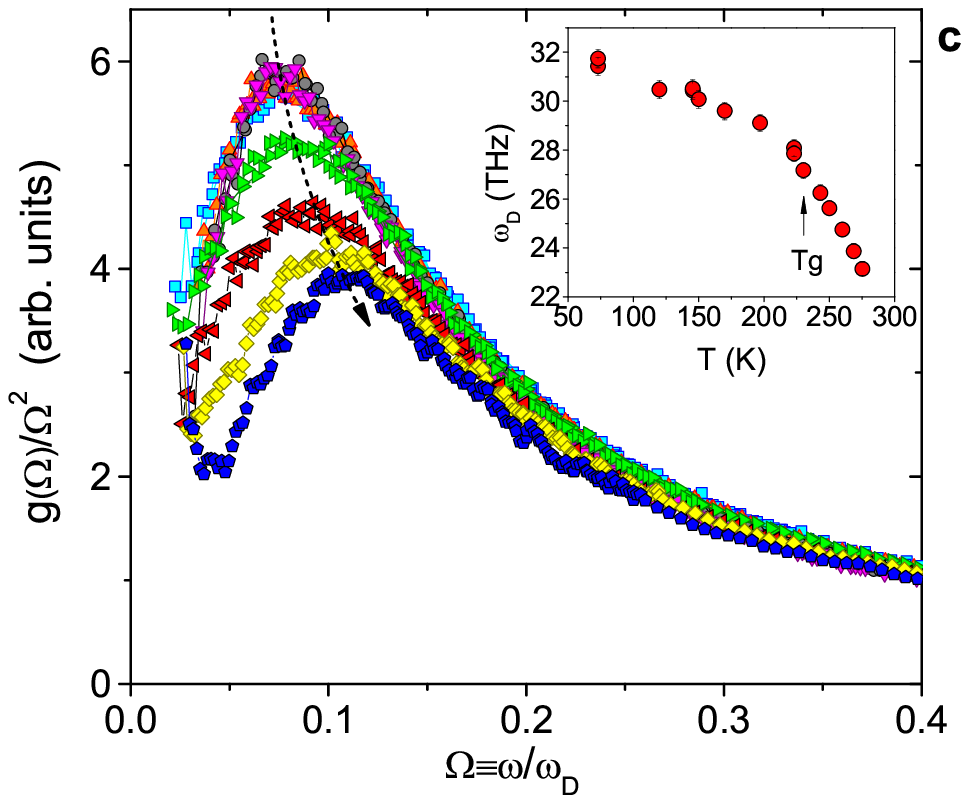}
\includegraphics[width=7.5 cm]{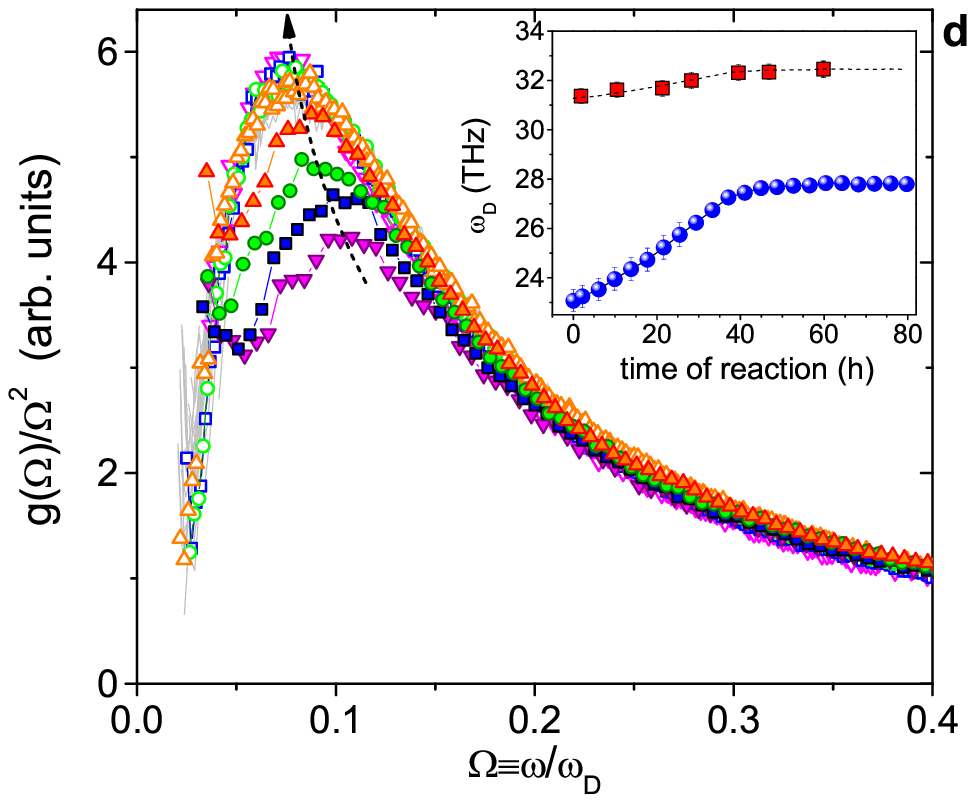}
\vspace{-0.8 cm} \caption{\label{Fig:temp} (a) The reduced density
of vibrational states, $g(\omega)/\omega^{2}$, for the unreacted
DGEBA-DETA 5:2 mixture at different temperatures as indicated in the
legend. (b) The quantity $g(\omega)/\omega^{2}$ for the DGEBA-DETA
5:2 mixture at different reaction times during the isothermal
polymerization at 275 K (data from ref.~\onlinecite{CaponiPRL2009}
--- grey lines and open symbols), and after quenching at 73 K some
partially polymerized states (solid symbols) as indicated in the
legend. (c) Reduced density of vibrational states after rescaling by
the Debye frequency $\omega _{D}$. Symbols are the same as in panel
(a). Inset: $\omega _{D}$ as function of the temperature. (d)
Reduced density of vibrational states after rescaling by the Debye
frequency $\omega _{D}$. Symbols are the same as in panel (b).
Inset: $\omega _{D}$ as function of the reaction time, during the
isothermal polymerization at 275 K ($\bullet$) and after quenching
at 73 K seven partially polymerized states ($\blacksquare$).}
\end{figure*}

It is well known that by changing the thermal history or the
thermodynamic path to the glass transition the obtained glasses
present different elastic properties, both in term of sound velocity
and in term of acoustic attenuation \cite{LevelutPRB2006,
CaponiPRB2004}, because of the different amount of internal stresses
that develop for the inability of the system to comply with the need
to contract under external stimuli \cite{WithersRPP2007, Tesi,
Kruger2003, BallauffPRL2013}. The stress-induced modification of
elastic properties in the glass can influence the vibrational
dynamics and may have a role in the evolution of the BP. So far this
issue has never been considered but here we clearly demonstrate, for
the first time, the occurrence of a stress-induced modification in
the BP scaling behavior of glasses. Starting from a reactive mixture
in the fluid phase, and combining the system's ability to achieve
the glassy phase both by decreasing the temperature (thermal
vitrification) and by increasing the number of bonds among the
constituent monomers (chemical vitrification
\cite{CorezziNATURE2002, CorezziPRL2005, CorezziPRL2006}), we
generate glasses with progressively different, either increasing or
decreasing, residual stresses. The introduction of significant
levels of internal stress in the system leads to a clear departure
of the elastic moduli from the generalized Cauchy relation. Our
results from Raman (RS) and Brillouin light scattering (BLS) provide
evidence that the scaling of the BP with the properties of the
elastic medium --- as measured by the Debye frequency --- breaks
down when the Cauchy-like relationship starts failing, and tends to
be recovered when the amount of residual stress is reduced.

\section{Experiments}

Our system is an epoxy-amine mixture of diglycidyl ether of
bisphenol-A (DGEBA) and diethylenetriamine (DETA) in the 5:2
stoichiometric ratio. The two monomer types are mutually reactive
and polymerize by stepwise addition with a rate of reaction strongly
controlled by the temperature. At $T$=275 K the reaction runs out in
$\sim 2.5$ days. The polymerization proceeds until the particles
diffuse and the unreacted sites can become close to each other, then
stopping when the average particle diffusion decreases to vanishing
levels \cite{CorezziJPCB2010, CorezziPOLYMER2010}. Therefore, the
system spontaneously evolves into a glassy structure (in the reacted
mixture $T_{g}^{reacted}\sim 300$ K). As shown in ref.
\onlinecite{CaponiPRL2009}, throughout the isothermal reaction the
BP scaling law remains Debye-controlled and the elastic moduli
follow (within the errors) the Cauchy-like relation. At any time
during the polymerization, however, the system
--- as a usual glass former --- can also be forced to reach the
glassy phase by decreasing the temperature. Exploiting this
versatility, we have designed the following two experiments to push
the system into states with different amounts of residual stress:
(i) The first is a cooling experiment of the unreacted mixture from
above to well below the glass transition ($T_{g}^{unreacted}$=231
K). The freshly prepared mixture is progressively cooled from 275 K
to 80 K, each temperature step is performed at $\sim 2$ K
min$^{-1}$. Since the rate of reaction at lower temperature is
further reduced, the advancement of reaction is prevented and only
the thermal vitrification route remains viable. The system is always
in equilibrium above $T_{g}$, while the chosen cooling rate is fast
enough to induce an \emph{increasing} amount of unrelaxed stress in
the glassy structure when the temperature is decreased below $T_{g}$
\cite{Kruger2003}. (ii) The second is a quench experiment performed
during the isothermal polymerization of the mixture. At different
times as reaction proceeds at 275 K, the partially polymerized
mixture is quenched to 73 K at $\sim 2$ K min$^{-1}$ and then
brought back to the reaction $T$. The effect of an isothermal
polymerization is combined with the effect of cooling in order to
obtain at the same $T$ glasses with higher polymerization level and
\emph{decreasing} amount of unrelaxed stress.

\begin{figure}
\vspace{-0.3 cm}
\includegraphics[width=8. cm]{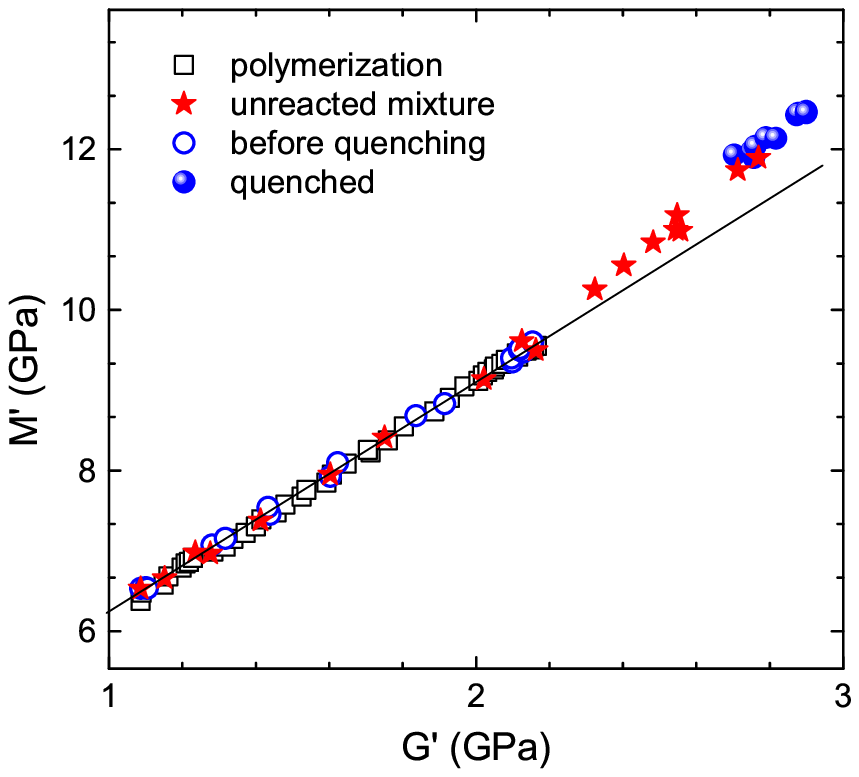}
\vspace{-0.5 cm} \caption {\label{Fig:Cauchy} Real part of the
longitudinal modulus $M'$ vs. real part of the transverse modulus
$G'$ of the DGEBA-DETA 5:2 mixture during the isothermal
polymerization reaction at 275 K ($\square$) \cite{Cauchy}, under
cooling the unreacted mixture ($\bigstar$), and after quenching at
73 K the sample at different extents of polymerization ($\bullet$).
Open circles represent the elastic moduli in the partially
polymerized states at 275 K before quenching. Error bars are
comparable with the symbol size. The solid line is the Cauchy--like
relation $M' = (3.16 \pm 0.08) + (2.99 \pm 0.02) G'$.}
\end{figure}

In all experiments, the two reagents were mixed and stirred for
about 2 min, and then transferred in the measurement cell (a
cylindrical pyrex vial of inner diameter 10 mm). Raman spectra were
acquired in the frequency range $3-1700$ cm$^{-1}$, using a Jobin
Yvon U1000 in the HV and VV polarization geometries, where V and H
indicate the polarization perpendicular and parallel to the
scattering plane. At temperatures lower than $\sim$ 220 K the
subtraction of a luminescence background was needed to restore the
symmetry of the Stokes and anti-Stokes intensities after reduction
by $[n(\omega)+1]\omega$, where $n(\omega)+1$ is the Bose population
factor. The reduced spectra were then normalized to the intensity of
the molecular vibration peaks at high frequency ($\omega > 600$
cm$^{-1}$). The depolarization ratio,
$\rho(\omega)=I_{HV}(\omega)/I_{VV}(\omega)$, has the constant value
$0.75\pm 0.02$ in the frequency range 5-100 cm$^{-1}$, independent
of temperature and consistent with the value $0.77\pm 0.03$ that we
measured during the polymerization process. The depolarized Raman
intensity $I_{HV}(\omega)$ has been used to obtain a quantity
proportional to the reduced density of vibrational states
$g(\omega)/\omega^{2}$, following the same procedure as in
ref.~\onlinecite{CaponiPRL2009}. In particular, the quasielastic
scattering contribution $I_{QES}(\omega)$ has been described with a
Lorentzian tail \cite{YannopoulosPRE2002}, subtracted from the total
signal, and the remaining vibrational component $I_{BP}(\omega)$ has
been related to $g(\omega)$ through the effective relation
$I_{BP}(\omega)=C(\omega)g(\omega)[n(\omega)+1]/ \omega$, where
$C(\omega)$ is the Raman coupling function that we reasonably
approximate as $C(\omega) \propto \omega$ for frequencies around the
BP maximum \cite{SokolovPRB1995, SurovtsevPRB2002, SurovtsevPRB2003,
FontanaJPCM2007, SchulteVIB2008} and assume not to change
appreciably with $T$ \cite{CaponiPRB2007, SokolovPRB1995}.

Brillouin spectra were acquired in the $90^{\circ}$--scattering
geometry using a Sandercock-type 3+3-pass tandem Fabry-P\'{e}rot
interferometer, with $\lambda$=532 nm and no selection of the
polarization for the scattered light (VU). This configuration allows
to detect both longitudinal acoustic modes, which scatter light
without change of polarization, and transverse acoustic modes, which
scatter light with change of polarization. The corresponding sound
velocities have been calculated as $v_{L,T}= \nu_{L,T} \lambda/(2 n
\sin\theta/2)$, where $\nu_{L,T}$ are the frequencies of the
acoustic modes, $n$ the refractive index of the sample, and $\theta$
the scattering angle. The resulting values have been processed along
with the results of previous BLS and IXS experiments
\cite{CaponiPRL2009}, in order to obtain the longitudinal and
transverse sound velocities in the high-frequency, solid-like limit
(details are given in the Appendix).

\section{Results}

Fig.~\ref{Fig:temp}(a) shows the obtained $g(\omega)/ \omega^{2}$
for the unreacted mixture in a wide temperature range, above and
below the glass transition. The BP evolution is clearly visible: as
the temperature decreases, the BP shifts towards higher frequencies
and decreases in intensity. The change is of $\sim 100\%$ in
frequency position and $\sim 400\%$ in height when $T$ decreases
from 275 to 80 K.

Fig.~\ref{Fig:temp}(b) shows the same quantity, $g(\omega)/
\omega^{2}$, measured during the polymerization at 275 K, and after
quenching at 73 K the sample at different extents of polymerization.
On increasing the number of covalent bonds under constant $T$ and
$P$ the system changes from liquid to glass and the BP shows
qualitatively similar changes to those observed on cooling the
unreacted mixture. However, the BP evolution of the quenched states
reveals a non-trivial behavior. In fact, each quench produces, as
expected, an additional moving of the BP position toward higher
frequencies and a  significant decrease in height, but this effect
has not the same entity at different extents of polymerization.
Indeed, the first quench, performed after 1 hour of reaction,
doubles the frequency of the BP maximum and reduces of $\sim 3.5$
times its intensity; the last quench, performed on the fully
polymerized mixture after 60 hours of reaction, only increases of
$\sim 30\%$ the frequency of the BP maximum and reduces of $\sim
1.6$ times its intensity. As a result, the BP measured at 73 K at
increasing extents of polymerization shifts to lower frequencies and
increases in height, clearly in contrast to the evolution that would
be expected in \emph{equilibrium} at fixed $T$. In the following, we
will consider these variations of the BP in relation to the
transformation of the elastic medium.

To quantify the elastic medium transformation, we calculate the
variations of the Debye frequency:
\begin{equation} \omega _{D}=(6 \pi^{2} \rho N_{A}N_{F}/M)^{1/3}\langle
v\rangle \label{eq:DebyeEnergy}
\end{equation}
where $N_{A}$ is the Avogadro's number, $N_{F}$ the average number
of atoms per molecule in the sample, $M$ the average molar weight,
and $\rho$ is the density, obtained using the procedure described in
ref.~\cite{Cauchy}. $\langle v\rangle$ is the Debye sound velocity,
defined as $\langle v \rangle ^{-3} = [(v_{L}^{\infty})^{-3}+2
(v_{T}^{\infty})^{-3}]/3$, where $v_{L}^{\infty}$ and
$v_{T}^{\infty}$ are the solid-like longitudinal and transverse
sound velocities, respectively \cite{Baldi2009, CaponiPRL2009}. It
should be emphasized that when the system undergoes a vitrification
process, whether physical or chemical, the sound velocities become
visco-elastic properties and acquire an intrinsic frequency
dependence. While an inelastic x-ray scattering (IXS) experiment,
performed in the THz frequency region, always probes
$v_{L}^{\infty}$, the sound velocities measured by BLS in the GHz
frequency region are, in the fluid phase, still affected by
viscoelastic effects and must be processed adequately in order to
obtain $v_{L}^{\infty}$ and $v_{T}^{\infty}$ (see details in the
Appendix). Using the density data and the proper values of sound
velocities, we calculate the Debye frequencies $\omega _{D}$
reported in the inset of Fig.~\ref{Fig:temp}(c) and
~\ref{Fig:temp}(d).

\section{Discussion}

Figs.~\ref{Fig:temp}(c) and ~\ref{Fig:temp}(d) show the evolution of
the reduced density of vibrational states corrected for the elastic
medium transformation, i.e. rescaled in Debye frequency units. It
appears a more complex behavior than that so far reported in the
literature, including regions of Debye scaling and regions of
non-Debye scaling of the BP. The significance of this behavior
emerges in connection with the behavior of the elastic moduli, shown
in Fig.~\ref{Fig:Cauchy}. The real part of the longitudinal and
transverse  modulus is measured by BLS as $M'=\rho v_{L}^2$ and
$G'=\rho v_{T}^2$, where $v_{L}$ and $v_{T}$ are the longitudinal
and transverse sound velocities at the BLS frequency. It is known
that the elastic properties of an \emph{isotropic} amorphous system
at \emph{equilibrium} satisfy a generalized Cauchy-like relation,
$M'=A + 3 G'$ \cite{Cauchy, KrugerPRB2002, PhilippJPCM2009}. Such
relation is theoretically understood for any isotropic material in
which particles interact by means of two-body central forces
\cite{ZwanzigJCP1965}. It is known as well, on experimental basis,
that stresses not released within an out-of-equilibrium glassy
matrix not only induce altered values of the elastic moduli
\cite{Tesi, WithersRPP2007} but also change the relationship between
them, leading to violation of the generalized Cauchy-like relation
\cite{Kruger2003}. Although not yet theoretically understood, this
behavior could be connected with the breakdown of ergodic assumption
in the glassy state, and with the strong possibility that unrelaxed
macroscopic stresses develop anisotropy in the material. As shown in
Fig.~\ref{Fig:Cauchy}, the relation $M' = (3.16 \pm 0.08) + (2.99
\pm 0.02) G'$ that connects in our system the elastic moduli during
the polymerization at 275 K also connects the moduli of the
unreacted mixture for temperatures above $T_{g}$. This linear
relation therefore characterizes, independently from the stiffening
path, the states in which the system is able to release internal
stresses and is our tool to discriminate them from the stressed
ones. As expected, the data clearly confirm the presence of residual
stress in the glassy structure of the unreacted mixture below
$T_{g}$ and in all the quenched states of the partially polymerized
sample. In particular, the data for the unreacted mixture deviate
from the Cauchy--like behavior by following a linear trend with
slope higher than 3, indicating the onset of internal stress just
below $T_{g}$.

With this in mind, the behaviors in Figs.~\ref{Fig:temp}(c) and
~\ref{Fig:temp}(d) can be read as follows. (i) In the cooling
experiment, the corrections for the elastic medium completely remove
the differences in the BP at temperatures above $T_{g}$, where the
system is able to release internal stresses. As $T$ decreases below
$T_{g}$, the system retains internal stresses, and correspondingly
the Debye scaling for the BP breaks down.  The influence of
temperature on the BP becomes stronger than the transformation of
the elastic medium, and specifically, the \emph{rescaled} spectra
progressively move to higher frequency and decrease in intensity
[see the arrow in Fig.~\ref{Fig:temp}(c)], thus indicating an actual
progressive loss of vibrational states compared to the system at
equilibrium.  (ii) In the quench experiment, internal stress is
accumulated in the sample and the Debye-scaling properties that
characterize both the liquid and the glassy phase of the isothermal
polymerization get lost. Again, the BP in Debye units stays higher
in frequency but lower in intensity compared to the master curve at
equilibrium, confirming a deficit of vibrational states in the
presence of a significant amount of internal stresses. However, it
should be noted that in this case the \emph{rescaled} spectra have
an opposite trend [see the arrow in Fig.~\ref{Fig:temp}(d)]: they
move to lower frequency and increase in intensity, suggesting an
eventual match with the master curve generated in a stress-free
condition. This indicates that reducing internal stresses tends to
restore the Debye scaling of the BP.

\begin{figure}
\includegraphics[width=7.5 cm]{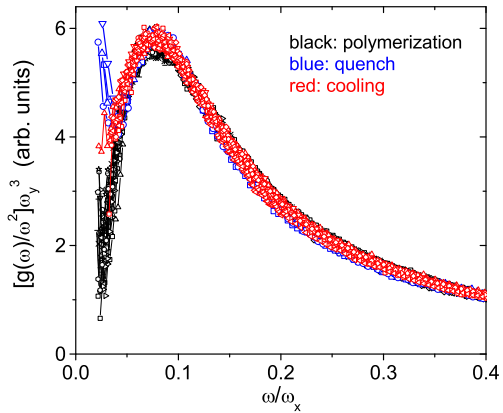}
\vspace{-0.5 cm} \caption {\label{Fig:superscaling} Master curve
obtained from $g(\omega)/\omega^{2}$ shown in
Figs.~\ref{Fig:temp}(a) and \ref{Fig:temp}(b), by plotting
$[g(\omega)/\omega^{2}]\omega_{y}^{3}$ vs. $\omega/\omega_{x}$. The
spectra refer to the DGEBA-DETA mixture at different reaction times
during isothermal polymerization at 275 K (black symbols), the
unreacted mixture at different temperatures (red symbols), and the
mixture after quenching at 73 K some partially polymerized states
(blu symbols). The spectra acquired during polymerization and at
temperatures above $T_{g}$ are scaled by using the measured Debye
frequency, i.e. $\omega _{x}=\omega _{y}=\omega _{D}$. The spectra
acquired in the quenched states and at temperatures below $T_{g}$
are scaled by using appropriate factors $\omega _{x}\neq\omega
_{y}\neq\omega _{D}$. Notice that the BP retains the same scaled
spectral shape.}
\end{figure}

It is interesting to note that independently from the validity of
the Debye scaling the spectral shape of the BP remains unchanged
(Fig.~\ref{Fig:superscaling}). When the spectra do not scale with
$\omega _{D}$ --- i.e., when the representation
$[g(\omega)/\omega^{2}]\omega_{D}^{3}$ vs. $\omega/\omega_{D}$ is no
longer generating a master curve --- the spectra can still be scaled
on each other using different scaling factors for the BP position
and intensity. This indicates that changes in the distribution of
modes around the BP occur in a way that the scaled spectral shape
remains essentially unaffected, despite the system undergoes strong
changes of physical and chemical nature. Our case broadens the class
of materials where invariance of the BP spectral shape has been
observed \cite{MonacoPRL2006hyp, MonacoPRL2006den, Baldi2009,
Ruffle2010, Zanatta2011, Zanatta2010, CaponiPRL2009, NissPRL2007,
Hong2008, CaponiPRB2007}, and offers an example of invariance even
against the formation of chemical bonds between the constituent
molecules, suggesting that the BP distribution is a system's
characteristic.

\begin{figure}
\includegraphics[width=7.5 cm]{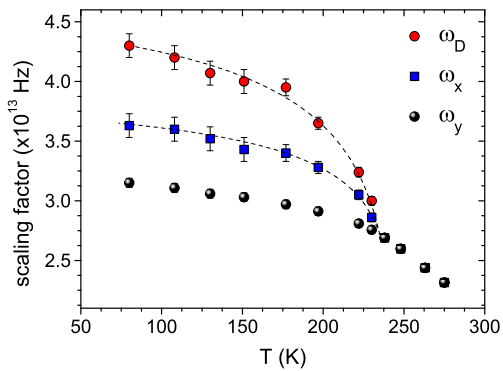}
\vspace{-0.5 cm} \caption {\label{Fig:scalingfactor} Debye
frequency, $\omega _{D}$, compared with the frequencies $\omega
_{x}$ and $\omega _{y}$ used to scale $g(\omega)/\omega^{2}$ in
Fig.~\ref{Fig:superscaling} for the unreacted mixture as a function
of $T$. Dashed lines are only guides for the eyes.}
\end{figure}

In the case of non-Debye scaling, by denoting with $\omega _{x}$ and
$\omega _{y}^{-3}$ the factors that scale, respectively, the BP
position and intensity on the master curve generated in a
stress-free condition, we always find in our system $\omega _{D}<
\omega _{y}< \omega _{x}$. This result is shown in
Fig.~\ref{Fig:scalingfactor} for the unreacted mixture as a function
of $T$. We remark that it is in agreement with the behavior of the
BP position and intensity observed in silica as a function of $T$,
considering that silica becomes stiffer (higher elastic moduli) when
$T$ increases \cite{Ruffle2010}. On the contrary, studies of BP as a
function of pressure \cite{NissPRL2007, AndrikopoulosJNCS2006}
report an increase with $P$ in the BP intensity relative to the
Debye level ($\omega _{y}< \omega _{D}$) while the BP frequency in
Debye units increases ($\omega _{x}> \omega _{D}$). The difference
in the intensity behavior could indicate a difference in the effect
of $T$ and $P$ on the vibrational modes in the glass under stress,
which in the first case experiences negative pressure rather than
positive. However, additional experimental evidences are needed to
assess the generality of this observation.

\begin{figure}
\includegraphics[width=8. cm]{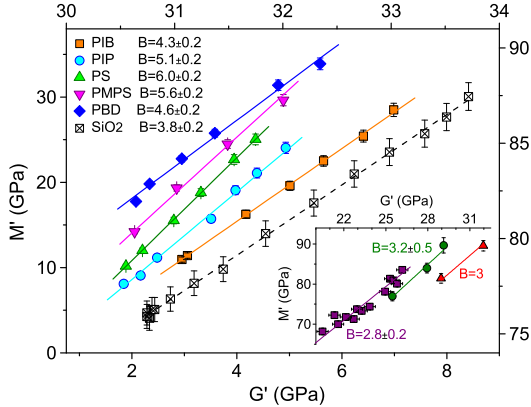}
\vspace{-0.5 cm} \caption {\label{Fig:CauchyNO} Longitudinal modulus
$M'$ vs. transverse modulus $G'$ of vitreous silica under cooling
($\boxtimes$; data from ref.~\cite{Ruffle2010}, to be read on the
right and top axes), and of five polymers under pressure (see
legend; data from ref.~\cite{Hong2008}). For clarity, the data of
PS, PMPS and PBD are vertically shifted by +2, +6 and +9,
respectively. The lines are a linear fit, $M' = A + B G'$. Inset:
$M'$ vs. $G'$ in sodium silicate glasses --- cooled ($\square$; data
from ref.~\cite{Baldi2009}); hyperquenched ($\bigtriangleup$; data
from ref.~\cite{MonacoPRL2006hyp}); permanently densified
($\bigcirc$; data from ref.~\cite{MonacoPRL2006den}).}
\end{figure}

All together our results can help provide a rationale for the
apparently conflicting conclusions of previous investigations of BP
scaling behavior. The key point is that existing studies are limited
to glass-formers below their glass transition and hence always in a
condition in which the particle configurations cannot fully relax to
equilibrium and the system cannot fully release internal stresses.
Therefore, they should not give a Debye-scaled BP. However, on the
basis of our results, if the amount of stress retained within the
glassy structure is small, such that deviations from the
Chauchy-like relationship cannot be appreciated within the
experimental uncertainty, then the BP is expected to scale
reasonably well with the elastic properties of the medium; otherwise
one expects the BP scaling does not work. To test this expectation,
we check the validity of the Cauchy--like relation in all the so far
studied systems (Fig.~\ref{Fig:CauchyNO}). Available $M'$ and $G'$
data for systems in which the BP evolution is not Debye-controlled
are collected in the main panel of Fig.~\ref{Fig:CauchyNO}, while
data for systems in which the BP reasonably scales with $\omega_{D}$
are presented in the inset. Surprisingly, all data in the main panel
exhibit an angular coefficient ($B$) significantly higher than 3
revealing the presence of stress in the investigated samples, while
the data in inset are all compatible with a low-stress condition
($B\approx 3$) \cite{nota}. In this regard, our study suggests that
if internal stress would be drastically reduced by applying
appropriate annealing treatments to the stressed samples, then the
Debye scaling of the BP would tend to be recovered.

\section{Conclusions}

By a combination of experiments performed during polymerization and
quenching of a reactive mixture, we provide evidence that the Debye
scaling of the BP holds for states in which the system is able to
release internal stresses, whereas deviations from the scaling
originate in the presence of residual stress, evidenced by
deviations from the generalized Cauchy relation. In particular, we
have shown that for our system, independently from the vitrification
path, the BP spectra can be scaled into a single master curve by
means of two independent parameters, one for the frequency axis and
one for the intensity; in case the residual stresses are negligible
these parameters are equal and coincide with the value of the Debye
frequency. Therefore, while the scaled spectral shape of the BP
appears to be a system's characteristic, there may be different
scaling regimes controlled by the presence of residual stress. The
reported correlation between breakdown of the Debye scaling and
violation of the Cauchy-like relationship is also able to give a
rationale of all the previous conflicting results in the literature.

\begin{acknowledgments}
We thank M. Mattarelli for the helpful discussions.
\end{acknowledgments}

\appendix*
\section{Derivation of solid-like sound velocities}

\begin{figure}
\includegraphics[width=9 cm]{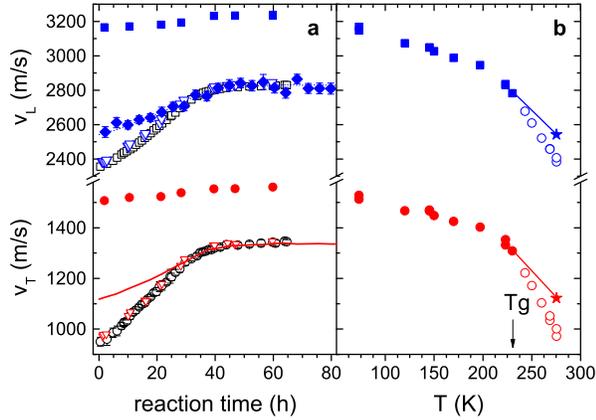}
\vspace{-0.3 cm} \caption {\label{Fig:sound} (a) Longitudinal and
transverse ($\square$, $\circ$) sound velocity measured by BLS
during isothermal polymerization at 275 K, \cite{CaponiPRL2009} and
after quenching at 73 K some partially polymerized states
($\blacksquare$, $\bullet$). The velocities at 275 K measured just
before and after the quench procedure are indicated with triangles.
The BLS values during polymerization are compared with their
high-frequency values $v_{L}^{\infty}$ ($\blacklozenge$) and
$v_{T}^{\infty}$ (solid line), which are obtained by IXS
\cite{CaponiPRL2009} and using the Cauchy-like relation.
\cite{Cauchy} (b) Longitudinal ($\square$, $\blacksquare$) and
transverse ($\circ$, $\bullet$) sound velocity measured by BLS in
the unreacted mixture as a function of temperature. A star indicates
the longitudinal high-frequency value $v_{L}^{\infty}$ measured by
IXS at 275 K. The solid line above is the linear interpolation of
this value to the value measured by BLS at $\sim T_{g}$, and the
solid line below is the high-frequency transverse sound velocity
estimated from the longitudinal one using the Cauchy-like relation.}
\end{figure}

BLS and IXS measurements as a function of time during isothermal
polymerization of the DGEBA-DETA reactive mixture at 275 K were
previously reported; \cite{CaponiPRL2009, CorezziPRL2006} in the
present work, BLS measurements have been extended as a function of
temperature by cooling the unreacted mixture from 275 K to 80 K, and
after different quenches (to 73 K) of the mixture at different times
of polymerization. From these experiments, we have derived the
appropriate solid-like velocities $v_{L}^{\infty}$ and
$v_{T}^{\infty}$ that are needed to calculate $\omega _{D}$ in all
the experimental conditions of the present study, as follows:

(i) During the isothermal polymerization at 275 K, $v_{L}^{\infty}$
is directly measured by IXS and $v_{T}^{\infty}$, which is not
accessible in the IXS experiment, is obtained using the Cauchy-like
relation $M' = (3.16 \pm 0.08) + (2.99 \pm 0.02) G'$, that connects
the longitudinal modulus $M'=\rho v_{L}^2$ and the transverse
modulus $G'=\rho v_{T}^2$ at all frequencies, with $\rho$ the
density \cite{Cauchy}. The $v_{T}^{\infty}$ data as function of the
polymerization time are shown in Fig.~\ref{Fig:sound}(a) with a
solid line.

(ii) In the quenched states at 73 K the system is well into the
glassy phase and BLS directly measures the solid-like elastic values
of both $v_{L}^{\infty}$ and $v_{T}^{\infty}$. The data are reported
as solid circles and squares in Fig.~\ref{Fig:sound}(a).

(iii) For the unreacted mixture under cooling, in the temperature
region below $T_{g}$ the sound velocities measured by BLS already
correspond to the high-frequency values $v_{L}^{\infty}$ and
$v_{T}^{\infty}$ (solid symbols in Fig.~\ref{Fig:sound}(b)), while
in the temperature region above $T_{g}$ the IXS measurement made at
the beginning of the isothermal polymerization provides the value of
$v_{L}^{\infty}$ at 275 K (this value is marked with a star in
Fig.~\ref{Fig:sound}(b)). Joining this last point to the point at
$T\sim T_{g}$, by assuming a linear dependence on temperature as
observed for the BLS data (open symbols), allows us to obtain
$v_{L}^{\infty}$ also in the viscoelastic fluid. The corresponding
values of $v_{T}^{\infty}$ are obtained from the Cauchy-like
relation \cite{Cauchy}; these results are reported in
Fig.~\ref{Fig:sound}(b) with a solid line.

\begin{references}
\bibitem{nota} We notice that the fairly good scaling of the BP reported in ref. \cite{MonacoPRL2006hyp}
for the hyperquenched and annealed glass, and the apparent agreement
with the Chauchy-like relation are quite surprising. In this case,
it is possible that a small deviation from the Debye scaling does
exist, although hardly discernible from the experimental error, and
also some deviation from the Cauchy-like relation might arise from
more data points (note that only the data for one hyperquenched
sample are available). We can only conclude that this case is
compatible with a low-stress condition, possibly favored by the
small dimensionality of the sample. On the other hand, the system is
subjected to small changes of $\omega_D$ ($\sim 5\%$) and therefore
a large deviation from the scaling is unlikely.
\end{references}
\end{document}